\documentstyle[12pt]{article}

\begin{document}
\bibliographystyle{unsrt}
\textwidth 800pt

\large
\begin{center}
\underline{Localized excitations in two-dimensional
Hamiltonian lattices
}
\vspace{2cm}\\ \large S. \vspace{2cm}Flach$^1$, K. Kladko$^{2,*}$ and
C. R. Willis$^1$ \\
\end{center} \normalsize
$^1$Department of Physics, Boston University, 590 Commonwealth Avenue,\\
Boston, Massachusetts 02215, \small flach@buphy.bu.edu
\\
\\
$^2$ Department of Physics and Astronomy, University of Maine, \\
5709 Bennet Hall, Orono, ME 04469-5709.
\\
\\
$^*$ on leave of absence from: Department of Physics, Kharkov State
University,\\
Kharkov, 310077 Ukraine
\vspace{1cm}
\newline
\normalsize
ABSTRACT \\

We analyze the origin and features of localized excitations
in a discrete two-dimensional Hamiltonian lattice.
The lattice obeys discrete translational symmetry, and the localized
excitations exist because of the presence of nonlinearities.
We connect the presence of
these excitations with the existence of local integrability
of the original N degree of freedom system. On the basis
of this explanation we make several predictions about
the existence and stability of these excitations.
This work is an extension of previously published results
on vibrational localization in one-dimensional nonlinear
Hamiltonian lattices (Phys.Rev.E.49(1994)836). Thus we confirm
earlier suggestions about the generic property of Hamiltonian
lattices to exhibit localized excitations independent on the
dimensionality of the lattice.
\vspace{0.5cm}
\newline
PACS number(s): 03.20.+i ; 63.20.Pw ; 63.20.Ry
\newline
Submitted to: {\sl  Physical Review E
}
\newline
Date: 03/30/94

\newpage
\small

\section{Introduction}

In this contribution we will deal with vibrational localization
in Hamiltonian lattices {\sl without} any kind of disorder. We consider
a solution of
a set of coupled ordinary differential equations (CODE) of
an underlying Hamiltonian system. The localization property
of the solution implies the solution to be essentially zero (constant)
outside a certain finite volume of the system. Inside the specified
volume the solution has some oscillatory time dependence. The absence of
disorder implies the existence of certain
discrete (CODE) translational symmetries of all possible
solutions.

Usually vibrational localization can be produced by considering
a lattice with a defect (diagonal or off-diagonal disorder) \cite{hb83}.
Another wellknown possibility is to consider lattices with
more than one groundstates (global minima of the potential energy)
and static kink-like distortions of the lattice. The presence
of the kink-like static (stable) distortion of the lattice
breaks the discrete translational symmetry as in the case of
a defect. This is the key ingridient to get localized
vibrations (localized modes) centered around either the defect
or the kink-like distortion \cite{hb83}.
It is worthwhile to mention that the existence of kink-like
distortions implies the underlying Hamiltonian lattice
to be nonlinear.

However it was known for a long time that special
partial differential equations admit breather solutions.
These breather solutions are exact localized vibrational
modes, which require neither disorder nor kinks.
In the case of the sine-Gordon (sG) equation the {\sl tangent}
of the breather solution
is given by a product of a space-dependent and a periodic time
dependent functions \cite{sdk53}.
The sG system has a phonon band with
a nonzero lower phonon band edge (the upper band edge is
not present, since its finiteness would imply the
discreteness of the system). The fundamental frequency of
the breather lies in the phonon gap below the phonon band.
The representation of the inverse
tangent of the periodic
time master function in a Fourier series shows up with contributions
from higher harmonics of the fundamental frequency. These higher
harmonics will certainly lie in the phonon band. The stability
of the breather solution in such a partial differential equation
will depend on some orthogonality properties between the
breather (higher harmonics) and the extended solutions (phonons)
\cite{ekns84}.
The fulfilling of all these
orthogonality relations seems to be connected to the
fact that the sG equation is integrable, i.e. admits an infinite
number of conservation laws.
Thus it appears logical that the sG breather
solutions survive only under nongeneric perturbations of
the underlying Hamiltonian field density \cite{bb93},\cite{jd93}.
Indeed efforts to find breather solutions in partial differential
equations of the Klein-Gordon type (i.e. closely related to
the sG case) failed e.g. for the $\Phi^4$ equation \cite{sk87}. The
$\Phi^4$ equation is not integrable.

Consider now instead of a partial differential equation
a Hamiltonian lattice. It will have at least one groundstate.
Generically the expansion of the potential energy around the
groundstate yields in lowest order a harmonical system and thus phonons.
However the phonon band will now have a finite upper band edge.
Thus we can imagine to create a breather-like localized
state with its frequency either above the phonon band
or even in a nonzero gap below the phonon band. In the first
case there will never be resonances between any harmonics
of the time function governing the evolution of the discrete
breather and phonons. In the second case we can again avoid
resonances by proper choice of the fundamental frequency
and the requirement that the phonon band width is smaller
than the gap width. Hence we seem to loose the necessity
to satisfy an infinite number of orthogonality relations
as in the continuum case. That could mean in turn that
the existence of discrete breather solutions will not be
restricted to the subset of nongeneric
Hamiltonian lattices.

Indeed over the past 6 years there have been several reports
on the existence of discrete breathers in various one-dimensional
nonintegrable Hamiltonian lattices of the Fermi-Pasta-Ulam
type and the Klein-Gordon type
\cite{st88},\cite{jbp90},\cite{cp90},\cite{bp91},\cite{th91},\cite{st92},\cite{sps92},\cite{bks93}. Unsurprisingly one will find no rigorous
derivation of the discrete breather solution in those reports, it
would be better to say that numerical results and several approximate
analytical results strongly imply the existence of discrete
breathers in one-dimensional nonintegrable Hamiltonian lattices.

Recently a careful study of the above mentioned system classes
revealed a first understanding for the phenomenon of
discrete breathers in terms of phase space properties of the
underlying system \cite{fw3},\cite{fw2},\cite{fw45},\cite{fw6}.
We will call these discrete breathers {\sl Nonlinear Localized
Excitations} (NLEs).
It was shown that the
NLE solution  can be reproduced with
very high accuracy considering the dynamics of a reduced
problem.
In the reduced problem one keeps the few degrees
of freedom which are essentially involved in the NLE solution of the
extended system. It turned out that the NLE solutions correspond
to regular trajectories in the phase space of the reduced problem.
These regular trajectories belong to a certain compact subpart
of the phase space which can be called regular island. The NLE regular
island is separated by a separatrix from other regular islands which
correspond to extended states in the full system.
Trajectories of the reduced problem on the separatrix
itself as well as in a certain energy-dependent part of the phase space
surrounding it are chaotic because the full system as well as
the reduced problem are non integrable. The whole emerging picture
we will call the local integrability scenario.

As it follows from that scenario, single-frequency NLEs correspond
to the excitation of one main degree of freedom which can be
characterized by its action $J_1$ and frequency
$\omega_1=\partial H / \partial J_1$ \cite{fw2}.
Many frequency NLEs correspond
to the excitation of several secondary degrees of freedom which
can characterized by their actions $J_m$, $m=2,3$ and frequencies
$\omega_m=\partial H / \partial J_m$ \cite{fw2}. Stability of the NLEs
in the infinite lattice environment can be studied with the
help of mappings. A certain movability separatrix can be defined
by $\omega_3=0$. This separatrix separates the phase space into
stationary NLEs (i.e. the center of energy oscillates around a
given mean position) and movable NLEs (i.e. the center of energy
can travel through the lattice) \cite{fw6}.

On the basis of the local integrability scenario it was
recently possible to {\sl prove} the generic existence of NLE solutions
in a one-dimensional nonlinear lattice with {\sl arbitrary} number of
degrees of freedom per unit cell and {\sl arbitrary} (still finite)
interaction range \cite{fw9}. Moreover  for the first time a rigorous proof
was given
that periodic NLE solutions {\sl do exist} in a class of Fermi-Pasta-Ulam
lattices \cite{fw9}.

{}From the local integrability scenario it follows that there are
no principal hurdles in going over to higher lattice dimensions
(by that we refer to the topology of the interactions rather than
the number of degrees of freedom per unit cell). Indeed the NLEs
are described through {\sl local} properties of the phase space
of the lattice {\sl and} no topological requirements on the potential
energy are necessary to allow for NLE existence. This is very
different compared to the well known topologically induced kink solutions,
for which the one-dimensional lattice is an analytical
requirement. Only under very special constraints can one
discuss kink-like solutions in lattices with higher dimensions.
Thus the NLE existence occurs to be a {\sl generic} property of
a nonlinear Hamiltonian lattice. Indeed a few numerical studies on NLEs in
two-dimensional Fermi-Pasta-Ulam lattices showed that NLEs
exist there \cite{bkp90},\cite{ff93}.

The purpose of this contribution is to apply the successfull
local integrability picture from one-dimensional lattices
\cite{fw3}-\cite{fw6}
to two-dimensional lattices. We will show that we indeed again
find NLE solutions (which are somewhat richer in their properties
compared to the one-dimensional case)
which are quantitatively describable with a reduced problem.
We will show this by comparing the phase space properties of
the full lattice and the reduced problem. We present a stability
analysis of the NLEs as well as a scheme to account for NLE
properties. Finally we present arguments about the statistical
relevance of the NLEs in the considered lattices at finite temperatures.
Thus we are able to show the correctness of our general approach
to vibrational localization in nonlinear Hamiltonian lattices and
of viewing NLEs as generic solutions in nonlinear discrete systems.

The paper is organized as follows. In section II we introduce the
model and briefly review the properties of NLEs in one dimension.
In section III  examples of NLE solutions in two dimensions
are presented. Then we define the reduced problem for the
two-dimensional system,
its phase space structure is compared to the corresponding part of
the phase space of the whole lattice.
A stability analysis is described, and different
evolution scenarios of NLEs are explained.
Section IV is used for a discussion of the results.

\section{Model, solutions in one dimension}

We study the dynamics of lattices with one degree of freedom
per unit cell and nearest neighbour interaction. The general
Hamiltonian is given by
\begin{equation}
H = \sum_{\vec{R}} \frac{1}{2}P_{\vec{R}}^2 +
\sum_{\vec{R}} V(X_{\vec{R}}) + \frac{1}{2}
\sum_{\vec{R}} \sum_{nn}\Phi(X_{\vec{R}} - X_{\vec{R'}}) \;\;. \label{2-1}
\end{equation}
Here $P_{\vec{R}}$ and $X_{\vec{R}}$ are canonically conjugated
momentum and displacement of the particle in the unit cell
characterized by the $d$-dimensional lattice vector $\vec{R}$. The
$d$ components of $\vec{R}$ are multiples of the lattice constant
$a=1$. The interaction and on-site potentials $\Phi(z)$ and
$V(z)$ are defined through
\begin{eqnarray}
\Phi(z)=\sum_{n=2}^{\infty}\phi_n \frac{z^n}{n!} \;\;, \label{2-2} \\
V(z)=\sum_{n=2}^{\infty}v_n \frac{z^n}{n!}\;\;. \label{2-3}
\end{eqnarray}
The abbrevation $(nn)$ in \ref{2-1} means summation over all
nearest neighbour positions $\vec{R'}$
with respect to $\vec{R}$.
Hamilton's equations of motion for the model are
\begin{equation}
\dot{X}_{\vec{R}}=P_{\vec{R}} \;\;, \;\; \dot{P}_{\vec{R}} = -
\frac{\partial H}{\partial X_{\vec{R}}} \;\;. \label{2-4}
\end{equation}
Thus we exclude from our consideration cases with i) more than one
degrees of freedom per unit cell and ii) larger interaction range.
The reasons for that are pragmatic - it will become too hard at
the present stage to present a careful study for the excluded
cases. We mention the numerical investigations of one-dimensional
chains with two degrees of freedom per unit cell in
\cite{ma92},\cite{ats93} and
some qualitative thoughts in \cite{yak93} about long range interactions,
where no indications of a change of the NLE existence properties
are found.

Let us briefly review the results for NLE properties in the one-dimensional
case. They are reported for two major subclasses of \ref{2-1}-\ref{2-3} -
the Klein-Gordon lattices \cite{fw3},\cite{fw2},\cite{fw45}
and the Fermi-Pasta-Ulam lattices \cite{fw6}. In the
case of Klein-Gordon lattices one drops the nonlinearities in the
interaction $\phi_2 = C \neq 0$, $\phi_{n > 2} = 0$ and allows for
nonlinearities to appear in the onsite potentials. Examples are the
$\Phi^3$ model ($V(z)=1/2 z^2 + 1/3 z^3$), the $\Phi^4$ model
($V(z) = 1/4 (z^2 - 1)^2$, the sine-Gordon model ($V(z)= \cos(z)$).
In the case of Fermi-Pasta-Ulam lattices one drops the on-site potential
$V(z)=0$ and allows for nonlinearities to appear in the interaction $\Phi (z)$.
In a convinient notation we will referr to them as FPU$klm$ models, where
$k,l,m$ are positive integers indicating the corresponding nonvanishing
power coefficients in \ref{2-3}. The Klein-Gordon systems have
up with a nonzero lower phonon band edge frequency (if $v_2 \neq 0$)
whereas the Fermi-Pasta-Ulam systems have a zero lower phonon band
edge frequency. Consequently the FPU models exhibit total momentum
conservation and show up with a zero frequency Goldstone mode -
in contrast to the Klein-Gordon lattices. Stable periodic (in time) NLEs
can be created in nearly all cited systems with frequencies outside
the phonon band (below or above for the Klein-Gordon systems, above
only for the FPU systems). The lowest NLE energy is nonzero -
i.e. there is a gap in the density of states of NLEs for energies
lower than the threshold energy. There can be gaps at higher energies too,
depending on resonance conditions between the NLE frequency and
the phonon frequencies. To allow for stable NLEs with frequencies
below the phonon band for Klein-Gordon systems one has to require
that the phonon band width is smaller than the phonon gap width.
One can understand the existence of a gap in the NLE density of states
by an approximate method to account for the NLE frequency. It consists
out of constructing an effective nonlinear one-particle potential. The energy
of a particle moving in this effective potential is the NLE energy,
and the fundamental frequency of its oscillation is the NLE frequency.
For small amplitude oscillations (small energies) the frequency
will lie always inside the phonon band of the corresponding
lattice. Increasing the amplitude (energy) will change the frequency
because of the nonlinearity. Depending on the type of the
effective potential the frequency can decrease or increase. At a certain
value of the amplitude (energy) the frequency leaves the phonon band,
thus the NLE becomes a stable excitation. This is also a very simple
guide to the prediction of the existence/nonexistence of NLEs in
nonlinear lattices. There will be no stable NLEs allowed to
exist in systems with e.g. a zero lower phonon band edge and
an effective potential of the defocussing type, i.e. where the
frequency will always decay with increasing amplitude (energy).
Instructive examples are the Toda lattice and the FPU23 lattice.

Because of the localization character of the NLE solutions essentially
only a finite number of particles are involved in the motion. Thus
it is possible to define a {\sl reduced problem} \cite{fw2}. It consists
of defining a finite volume around the NLE center. All particles
inside the finite volume are involved in the NLE solution, particles
outside essentially should not be involved. There is a uncertainty
in the definition of the finite volume. It comes from the
fact that the NLE solutions are not compact, i.e. strictly speaking
they incorporate an infinite number of particles (degrees of freedom)
\cite{fw7}.
But a sharp exponential decay of the amplitudes starting from the
center of the NLE provides a good finite volume choice in many
cases. Since the finite volume (reduced problem) consists out of
a finite number of degrees of freedom, it becomes easier to
analyze its phase space properties. As it was shown in \cite{fw2},
there exist regular islands in the phase space of the reduced problem.
These regular islands are separated by stochastic layers (destroyed
regular motions on and near separatrices) from each other. The motion
in each of the regular islands appears to be confined to a torus
of corresponding dimension. Certain islands can be labeled NLE islands.
Periodic orbits (elliptic fixpoints in corresponding Poincare mappings)
from these NLE islands
appear to be (nearly) exactly the periodic NLE solutions from
the full system. The surprise came when it was shown that the
quasiperiodic orbits surrounding the periodic one correspond
to many-frequency NLEs in the full system \cite{fw3},\cite{fw2}.
Although a stability analysis
shows that these many frequency NLEs are strictly speaking unstable
(i.e. they can not exist for infinite times) \cite{fw7} it turned out that
their
energy radiation rate can be very weak, such that the lifetimes
of these many frequency NLEs can become several orders of magnitude
larger than the typical internal periods. In numerical experiments
more than five orders of magnitude were easily found \cite{fw2}.
The lifetime of the many frequency NLEs will increase to infinity
if one chooses quasiperiodic orbits which are closer and closer to
the periodic orbit (the periodic NLE).

Other regular islands did not yield NLEs in the full system. The same
can be said about the orbits in the stochastic layer. The reason
for that is the resonance of the fundamental frequencies in those
regular islands with the phonon frequencies. Motion in the
stochastic layer is chaotic, thus frequency spectra are continuous
rather than discrete. Consequently generically there is always overlap
with the phonon frequencies and thus strong energy loss of the
finite volume. We also mention interesting long-time evolutional
scenaria for many frequency NLEs as described in \cite{fw2}.

The clear correspondence between regular islands in the reduced problem
and NLE solutions in the full system allows for a deep understanding
of the NLE phenomenon on one side. On the other side it opens
possibilities to apply the apparatus of nonlinear dynamics to
explore NLE properties. That was done in \cite{fw6} to study
the movability properties of NLEs.

In the following we will apply the same procedure to characterize
NLE solutions in two-dimensional systems. The success of our
study will have several impacts. First it will be a proof of
the conjecture that the NLE existence is not a specific one-dimensional
solution as e.g. the kinks. This conjecture was formulated
on the basis of the local integrability picture \cite{fw2} as described above.
Thus we strengthen the whole local integrability picture. Secondly
by establishing NLE solutions in two dimensional lattices
undoubtly will increase the interest in the overall phenomenon
because of the variety of physical applications in contrast to
the one-dimensional case. Moreover by proving the conjecture
about the unimportance of the dimensionality of the lattice
with respect to the NLE occurence also three-dimensional
applications become of potential interest.

\section{The two-dimensional case}

\subsection{Model specification, numerical details}

As an example we choose the $\phi^4$ lattice in two dimensions, i.e.
$V(z)=1/4 (z^2-1)^2$, $\Phi(z)=1/2 C z^2$, $\vec{R}= (l,m)$
with $l,m = 0,\pm 1, \pm 2, ...$ (cf. \ref{2-1} - \ref{2-3}).
The two groundstates of the system are given by
$X_{\vec{R}}= \pm 1$. The model has a phase transition
at a finite temperature $T_c$ which is
of no further concern here since we are studying properties
of single excitations above the groundstate (i.e. because
of the localized character of the solutions at effectively
zero temperature). The parameter $C$ specifies the 'discreteness'
of the system, i.e. the ratio of the phonon band width to
the phonon gap width. Since we are interested in
vibrations localized on a few particles, it is reasonable to
compare the onsite potential energy ($V(z)$) to the spring
energy ($\Phi(z)$) of a given particle when it is displaced
relative to its nearest neighbours. As it was shown in \cite{fw2}
besides the interaction parameter $C$ the energy (per particle)
becomes a second significant parameter in order to choose a reasonable
ratio between the two components of the potential energy.
One can easily take over the results from \cite{fw2} if one rescales
the parameter $C$ there by multiplying it with 2 (because in the
cited one-dimensional case the coordination number was 2 compared
to 4 in the two-dimensional case). Thus a choice of $C=0.05$ turns
out to be a case of intermediate interaction for not too large
energies, i.e. the on-site potential energy is of the same order
as the interaction potential energy.

The dispersion relation for small amplitude phonons (small
amplitude oscillations around either groundstate) is given
by
\begin{equation}
\omega^2_{k_x,k_y}= 2 + 4 C \left( \sin ^2(\frac{\pi k_x}{N})
+ \sin ^2 ( \frac{\pi k_y}{N}) \right) \label{phonon}
\end{equation}
where $N$ is the length of one side of the squared lattice,
and $k_x$ and $k_y$ are two integers under the condition
$0 \leq k_x,k_y \leq (N-1)$.

In all numerical simulations
a Runge-Kutta method of 5th order with time step $\Delta t = 0.01$
was used. We compared our results to an independent code
where a Verlet algorithm with $\Delta t = 0.005$ was used and observed
no differences.
In the studies of one-dimensional systems
the simulation of an infinite system was replaced
with a finite chain of
such a length that the fastest phonons could not make a turn and
come back to the finite volume of the NLE excitation during the
simulation time. In two dimensions such a method would mean a squared
waste of computing time and ban us on parallel computers. However
there is another way to avoid recurrence of phonons which are radiated
from the NLE - to switch on a (reflectionless) friction outside a
given volume such that the radiated phonons will be captured
and eliminated.
The condition of reflectionlessness implies a gradual increase
of the friction with growing distance or in other words a large number
of collisions between phonons and friction applied lattice sites.
The friction is added to the right-hand side of \ref{2-4}
in the form $-\gamma_{\vec{R}} P_{\vec{R}}$.
In the case of a full system we work with a friction-free volume
of size $20 \times 20$ and an additional friction-applied
boundary of thickness $10$ particle distances on each side.
Thus the overall number of particles is $40 \times 40 = 1600$.
The friction is linearly increased
in the friction-applied walls from zero up to a maximum value of $\gamma_0$
at the boundary layer.
At the boundaries periodic boundary conditions are applied.

To proceed we have to optimize the maximum friction coefficient,
since for $\gamma_0=0$ or $\gamma_0=\infty$ the phonons are
completely transmitted or reflected respectively.
We simulate the linearized $\Phi^4$
lattice ($V(z)=z^2$, $\Phi(z)=1/2 C z^2$)
with an initial condition, where the central particle
is displaced by $\Delta X = 1$ from its groundstate position,
all other particles are held at their groundstate positions
and the velocities are zero. The corresponding initial energy
is $E=1.1$. We let the system evolve, and measure the energy
stored in the system $E(t)$ and the energy stored in the central
particle and its four neighbours $E_5(t)$ for $t=2000$. The result
is shown as a function of $\gamma_0$ in Fig.1. We find that for the choosen
geometry the optimum value for the maximum friction coefficient is
$\gamma_0 \approx 0.005$. The full time dependence of the
two energies  $E(t)$ and $E_5(t)$ using $\gamma_0=0.005$ are
shown in Fig.2. We see that after waiting times of $t \leq 2000$
the central particle and its four neighbours loose more than
99.9\% of their initial energy. In the following we will use
the thus choosen value for the maximum friction coefficient
$\gamma_0=0.005$ in all described simulations.

\subsection{NLE solutions}

Let us show a stable NLE solution. For that we prepare the following
initial condition: central particle at groundstate position,
nearest neighbours displaced to $X_{(nn)}=-1.01163$, velocity
of nearest neighbours $P_{(nn)}=0.0225$, the velocity of the
central particle is adjusted to the initial energy $E=0.3$,
all other particles are at their groundstate positions with
zero velocities.
To characterize the localization properties we use
the local discrete energy density
\begin{equation}
e_{\vec{R}} = \frac{1}{2}P^2_{\vec{R}} +
V(X_{\vec{R}}) + \frac{1}{2}\sum_{nn}\Phi(X_{\vec{R}} - X_{\vec{R'}})
\;\;\;. \label{3-1}
\end{equation}
Let us define the energy stored on five particles
(the central particle $\vec{R}=(0,0)$ and its four neighbours)
\begin{equation}
e_{5}= \sum_{\vec{R}'} e_{\vec{R'}} \;\;, \;\;
|\vec{R'}| \leq 1 \;\;. \label{3-2}
\end{equation}
In the insert in Fig.3 we show $e_{5}$ as a function of time for the
above given initial condition. Clearly we observe localization
of vibrational energy for extremely long times. One has to keep in mind
that the typical oscillation times are of the order
of $t_0 = 4$.  The stability property of the observed NLE is very similar
to examples from one-dimensional cases. The energy distribution
in the NLE solution after $t= 3000$
is shown in Fig.3. Essentially five particles are involved in
the NLE motion - a central particle and its four nearest neighbours.

Since we used symmetrical initial conditions essentially two degrees
of freedom are excited. To describe the NLE solution we construct
a {\sl reduced problem} in analogy to the one-dimensional problem.
The reduced problem consists out of the five particles which are
essentially involved in the NLE motion. The rest of the lattice
is held at its ground state position. Together with the consideration
of symmetric initial conditions we are left with the following
two-degree of freedom problem:
\begin{eqnarray}
\ddot{Q}= Q - Q^3 + 4C (q-Q) \;\;, \label{3-3} \\
\ddot{q} = q - q^3 + C(Q-q) - 3C(1+q) \;\;. \label{3-4}
\end{eqnarray}
Here $Q=X_{(0,0)}$ and $q=X_{(\pm 1, \pm 1)}$ are the coordinates
of the central and nearest neighbour particles respectively.
In the one-dimensional case it was shown that certain solutions
of the reduced problem correspond to NLE solutions in the
full system \cite{fw2},\cite{fw45}.

\subsection{The reduced problem}

Before we show that the same correspondence principle works
for the two-dimensional example in the present work,
we want to characterize the main features of the system
of equations \ref{3-3}-\ref{3-4}.

In Figs.4(a-d) we show Poincare mappings for the reduced problem
for energies $E=0.2/0.5/2.5/5$. As can be seen there the reduced
problem is not integrable since we find stochastic motion.
Thus the energy is the only integral of motion. However
we find islands of regular motion (regular islands) which are
separated from each other by stochastic layers. The topology of the
stochastic
layers indicates the topology of destroyed separatrices.
For small energies $E=0.2$ (Fig.4(a)) the thickness of the stochastic
layer is too small to be detected at all (in the presented
resolution) so that we find two regular islands which we label
with the numbers 1 and 2. The elliptic
fixed points of each regular island correspond to time-priodic
solutions of the reduced problem. Increasing the energy we
find a rather abrupt increase of the thickness of the stochastic
layer for $0.35 < E < 0.4$. Thus at $E=0.5$ (Fig.4(b)) we are
faced with effects of period doubling (increasing number of
regular islands) and a decrease of the size of the islands.
For $E=2.5$ (Fig.4(c)) nearly the whole available phase space is
filled with chaotic trajectories. However for higher energies
(here $E=5$ in Fig.4(d)) the size of the regular islands increases
again. In the limit $E \rightarrow \infty$ the reduced problem
becomes infinitely close to an integrable system of two
noninteracting quartic oscillators.

A proper characterization of the regular islands is
the frequency of their corresponding elliptic fixpoints.
In Fig.5 the fixed point frequencies of the main regular islands
are shown as a function of energy. For small energies the
frequencies of the fixed points of regular islands 1 and 2
become the eigenfrequencies of the linearized problem (around
the groundstate): $\omega^2= 2+2C$ for island 1 and
$\omega^2=2+6C$ for island 2, both of the frequencies are
in the phonon band of the infinite system \ref{phonon}.
{}From Fig.5 it follows that there is a nonzero lower energy
threshold above which the fixed point frequency from island 1
becomes nonresonant with the phonon band.

For of reasons discussed below we concentrate on island 1.
Its fixed point frequency we denote by $\omega_1$ (here the
index refers not to the island number but to the degree of freedom
excited in the island). Then several statements can be made
with respect to the secondary degrees of freedom which can
be excited (cf. torus intersection structure around fixed
point in island 1). Considering an infintesimally small excitation
of the second (symmetric) degree characterized by its
frequency $\omega_2$ one can show that in the limit
of zero energy the $\omega_2^2 = 2+6C$ (cf. Appendix).
If one lifts the symmetry of the initial conditions in the
reduced problem one has to consider the generalized reduced
problem
\begin{eqnarray}
\ddot{Q}= Q - Q^3 + \sum_{i=1}^4 C(q_i - Q) \;\;, \label{3-5} \\
\ddot{q}_i = q_i - q_i^3 - 3C(1+q_i) + C(Q-q_i) \;\;. \label{3-6}
\end{eqnarray}
Here $Q = X_{(0,0)}$ as in \ref{3-3}-\ref{3-4} and the four coordinates
$q_i$, $i=1,2,3,4$ denote the coordinates of the four nearest
neighbours of the central particle. Since we deal with five
degrees of freedom now we have to expect five (instead of two)
fundamental frequencies. System \ref{3-5}-\ref{3-6} has rotational
symmetry of order 4. It follows (cf. Appendix) that the three
new frequencies $\omega_3$,$\omega_4$,$\omega_5$ are equivalent
to each other in the limit of infinitely small assymetric perturbations
of the fixed point periodic solution. In the limit of zero energy
it follows $\omega_3^2=\omega_4^2=\omega_5^2= 2 + 4C$. Increasing the
energy from its lowest value leads to a decrease of all five
frequencies. The inequality $\omega_1 < \omega_{3,4,5} < \omega_2$
(which is true only for low energy values)
determines the sequence of the $\omega_i$ crossings of the
lower phonon band edge.

\subsection{The correspondence principle}

Let us show the connection between the reduced problem and
the NLE solutions of the full system. For that we plot
in Fig.5 the frequencies of (nearly) periodic NLEs as a function
of energy. We observe very good agreement with the data of
the fixed point frequency $\omega_1$ from island 1 of the reduced
problem. In fact one can check that the whole time-dependent periodic
NLE solution of the full system is very close to the corresponding
fixed point periodic solution from island 1 of the reduced problem.
Since the frequency $\omega_2$ of the symmetric perturbation
of the fixed point periodic NLE solution according to the
results from the reduced problem is in resonance with phonon
frequencies up to energy values of 1, we increase the energy
to $E=5$ (cf. Fig.4(d)) and perform a Poincare mapping for
the NLE solutions of the {\sl full system}. The result is shown in Fig.6
{together with the corresponding data from the reduced problem}
(cf. Fig.4(d)). The result is amazing - the torus intersections
are practically identically for the two frequency NLE solution
from the full system and the corresponding regular trajectories
from the regular island of the reduced problem. If one chooses
an initial condition in the full system that corresponds to
the chaotic trajectory in the reduced problem (Fig4(b))
then we find a quick decay of the energy excitation in the
full system as shown in Fig.7.

If the energy is low enough the frequency $\omega_2$ of the
symmetric perturbation of the periodic NLE will come into
the phonon band. Then we expect a loss of the energy part
stored in the corresponding second degree of freedom, leaving
the main degree of freedom essentially unaffected. To show
that we simply perform a Poincare mapping for the mentioned
case. The result is shown in Fig.8. Indeed instead of an intersection
line with a torus we find a spiral-like relaxation of the NLE
solution onto the periodic fixed point NLE. The fixed point periodic
NLE acts like a limit circle, although the whole system is
conservative.

\subsection{Effective potential}

Because of the smallness of the nearest neighbours amplitudes
compared to the amplitude of the central particle in a NLE solution,
we can try to account approximately for the motion of the
central particle assuming the nearest neighbours are at rest at
their groundstate positions. Then the central particle would move
in an effective potential
\begin{equation}
V_{eff}(z)= V(z)+2C(z+1)^2 \;\;. \label{3-7}
\end{equation}
The motion in this potential is periodic with an energy-dependent
(or amplitude-dependent) period $T_1=2\pi / \omega_1$. Since most
of the energy in the NLE solution is concentrated on the central
particle and its binding energy to the nearest neighbours,
it is reasonable to compare the results for the energy dependence
of $\omega_1$ for \ref{3-7} with the numerical result as given
in Fig.5. As can be seen in Fig.5, the overlap between the
result from the effective potential, the reduced problem and
the full system are very good. Thus we have a proper method for
predicting the behaviour of the main NLE frequency $\omega_1$
as a function of energy. This method is directly taken over from
the known results in the one-dimensional case.

\subsection{Stability properties}

As it was shown in \cite{fw3},\cite{fw2} for the one-dimensional case,
it is possible to carry out a stability analysis for
periodic NLEs (i.e. the fixed point solutions) with respect
to extended phonon-like perturbations. In fact the
procedure for the stability analysis in the two-dimensional case
is exactly the same. Thus we will highlight here only the
necessary parts of the steps one has to follow.
We assume we know an exact periodic NLE solution $X_{\vec{R}}(t)$.
Then we consider a slightly perturbed trajectory $\tilde{X}_{\vec{R}}
= X_{\vec{R}}(t) + \Delta_{\vec{R}}(t)$. Since the assumed
NLE solution is localized, it becomes infinitely small
for large distances from the NLE-center. This circumstance does
not pose a serious problem for the definition of the expression
'slightly perturbed'. One can just consider small
amplitude oscillations (phonons) around the groundstate of our system.
Then we have a well-defined small parameter determining the
weak nonlinear corrections to the linear equations. We take over
this definition of smallness to our problem. In the center of the NLE
the perturbation will thus be small compared to the NLE-contribution.
Far away from the center the perturbation can even become large
compared to the NLE-contribution, but it will be still small enough
to ensure the linearized equations work well. Then we can consider
small perturbations of the NLE solution which are extended.

In the next step we insert the perturbed ansatz into the
lattice equations of motion. Using the fact that the unperturbed
part is a solution of the equations of motion and linearizing
the equations with respect to the perturbation yields a set
of coupled differential equations with time-dependent (periodic)
coefficients. In analogy to \cite{fw2} we can define a map, the stability
of which
is equivalent to the nongrowing of the small perturbation
of the NLE solution. The sufficient condition for the stability
is that neither of any multiple of half the NLE frequency
is equal to a phonon frequency \ref{phonon}:
\begin{equation}
\frac{\omega_{k_x,k_y}}{\omega_1} \neq \frac{n}{2}\;\;,
\;\; n = 0,1,2,... \;\;. \label{3-8}
\end{equation}
As we see this result explains the existence of an energy threshold
(gap in the density of NLE states) for the NLE solutions.
Because the NLE frequencies according to the reduced problem
will always lie in the phonon band for small enough NLE energies,
the low energy NLEs are unstable against smallest perturbations. In the
one-dimensional case this statement was tested in the full system
using an entropy-like variable measuring the degree of
energy localization \cite{fw2}. On approaching the energy threshold ( predicted
by the results of the reduced problem together with the stability
analysis) from above the entropy drastically increases at the predicted
threshold value. It is still possible that low-energy NLEs exist
with a very small degree of localization and
with frequencies very close to the phonon band edge, but outside the
band itself. In the two-dimensional case considered here
we also observe a very sharp transition in the degree of
localization at the predicted energy threshold value. In fact it
becomes impossible to find a NLE solution with energies below
the threshold value. That indicates the tiny phase space part
at low energies which might be still occupied with weakly localized
states.

A more subtle problem is the {\sl internal} stability of periodic
NLEs. As it is known for several one-dimensional systems,
periodic NLE solutions can become internally unstable, i.e. a
weak symmetrybreaking perturbation of the periodic NLE will
transform the NLE solution into other existing periodic NLEs
of different parity or even into NLEs moving through the lattice
\cite{cp90},\cite{sps92},\cite{cku93}.
Currently it is unclear how to classify and find the different
possible periodic NLE solutions on a two-dimensional lattice. First
efforts to do so are reported in \cite{ff93}. We wish to emphasize that
the periodic NLE solutions reported in this paper are certainly not
the only ones allowed to exit in the underlying lattice. Thus we
can only make statements about the internal stability of the NLE solutions
considered in the present work. Using the results of the linearization
of the equations of motion around the periodic NLE solutions (Appendix)
we can trace the values of the squared secondary eigenfrequencies
$\omega_2^2,\omega_{3,4,5}^2$ and can report here that throughout
the considered cases all squared eigenfrequencies are positive. Consequently
the periodic NLE solutions discussed here are internally stable.

The results of the {\sl stability} analysis drawn above do not allow
us to conclude about the {\sl existence} of NLE solutions in a strict
general sense. As it was shown in \cite{fw7} for the one-dimensional
case, periodic NLEs do not exist if any multiple of
the NLE frequency resonates with phonon frequencies (this condition
corresponds to the cases of even integers $n$ in \ref{3-8}).
Also all multiple frequency NLEs are strictly speaking unstable,
since it is always possible to find combinations
of multiples of two or more frequencies (whose ratio is irrational)
resonating with phonon frequencies. It appears currently unclear
how to take over the methods used in \cite{fw7} for the one-dimensional
case, to the two-dimensional case in order to obtain existence
criteria for NLE solutions. However it can be expected that the
methodological problems do not alter the results obtained in the
one-dimensional case.

As it was shown in \cite{tks88} the decay of the periodic
NLE solutions far away from the NLE center can be well described
by a Green's function method, which yields exponential decay in
the amplitudes.

\section{Discussion}

In the present work we have shown, that it is possible
to take over the results on the existence and properties
of nonlinear localized excitations in nonlinear lattices
from lattice dimension one to lattice dimension two.
Thus several goals were achieved: i) the existence of
NLEs in two-dimensional lattices is verified;
ii) the theory developed for NLEs in one-dimensional
systems appears to be of validity independent
on the lattice dimension; iii) the power of the theoretical
framework to predict the existence of NLE solutions
in several one-dimensional lattices has been extended by
its correct prediction of the NLE existence in higher dimensional
lattices. Thus the NLE existence in three-dimensional lattices
can be considered as highly likely. There is at the present
no single reason supporting the nonexistence of NLEs due to
lattice dimensions.

Besides the novel analysis of the properties of the secondary
frequencies (cf. Appendix) the present work has also shown,
that the resonating of secondary frequencies with phonon frequencies
does not imply a shrinkage of the phase space part of the system
corresponding to NLE solutions. Indeed as long as the main frequency
$\omega_1$ stays outside the band, the choice of an initial condition
with excited secondary degrees of freedom will
still yield a NLE.
If the secondary frequencies are outside the phonon
band as well, the solution will be a (very weakly) decaying
multiple frequency NLE. If the secondary frequencies resonate
with the phonon band, the corresponding energy part stored in
the NLE is radiated away and the NLE 'collapses' onto its
periodic fixed point solution. This attractor-like behaviour
ensures, that there is still a finite phase space volume
around the fixed point periodic solution which corresponds
to NLEs even after extremely long waiting times.
Thus we have strong evidence for the statistical relevance
of NLE solutions in corresponding lattices at finite temperatures.
Indeed the only case when NLEs can become statistically
unimportant is when the main frequency $\omega_1$ resonates
with the phonon band.

We can conclude from our results on the energy radiation
of perturbed periodic NLEs and from the mentioned existence
proofs for periodic NLEs, that the results on radiation processes
accounted for in \cite{bam94} are wrong. In order to get
the leading order radiation of perturbed periodic NLEs one has to
linearize the phase space flow of the system {\sl around}
the unperturbed periodic NLE solution, and {\sl not}
around the groundstate of the system as it was done in \cite{bam94}.

Let us finally address the question: what are the physical
applications where one can expect NLEs to exist?
In the mathematical sense the answer is: when the main
frequency $\omega_1$ can be 'pulled out' of the phonon
band with increasing energy. To check the behaviour of the
main frequency we have to construct the effective potential.
Consider e.g. a monoatomic crystal. The pair potential of interaction
is usually
an assymetric potential around the stability position.
The effective potential can be constructed exactly as described
in the previous section. If the repelling part of the pair potential
goes nonlinearly enough, then there can be oscillations of
a particle in the effective potential with frequencies above
the phonon band. However one has also to check the minimum energy
(energy threshold) required for the NLE existence. As studies
for a particular class of crystals have shown, NLEs can not be excited
there thermally, because the melting point is too low. However
it could be still possible to excite the NLEs locally nonthermally
\cite{bs91}.
If we consider crystals with many atoms per unit
cell, we can expect at least the existence of phonon gaps
between acoustic and different optical zones. It is
a well-known approach to describe structural phase transitions
with the use of $\Phi^4$-like models, simulating the behaviour
of certain soft phonon modes essentially decoupled from
other nonsoftening modes \cite{bc81}. However
there might be too many
problems on this path to really make sure that $\Phi^4$-lattice
type NLEs can exist in such crystals.
Another way of thinking leads us to the fact that the
adding of a periodic external field (onsite potentials)
can produce a finite phonon gap eliminating the conservation
of mechanical momentum. Such situations are very likely in
the case of atomic monolayers on proper substrate surfaces.
If it becomes possible to choose such cases, where the phonon band width
becomes
small compared to the gap width, NLEs could exist.
With these arguments we did not intend to judge
the different physical situations where NLEs are likely or
unlikely to exist. It was the search strategy we had in mind.
It is one of the forthcoming tasks to provide a foundation for these ideas
in order to proceed in the question of applicability. Still
the mathematical result that NLEs are generic solutions of
nonlinear lattices \cite{fw9} serves as a powerful indicator of their
relevance in different physical realizations.
\\
\\
\\
\\
Acknowledgements
\\
\\
We thank J. Krumhansl, E.Olbrich for interesting discussions,
S. Takeno and A. J. Sievers for sending us their preprints
prior publication, and J. Denzler for helpful comments.
One of us (K.K.) thanks the University of Maine for
warm hospitality and support.
This work was supported in part (S.F.)
by the Deutsche Forschungsgemeinschaft
(Fl 200/1-1).

\newpage
\appendix

\section{}

Here we consider equations \ref{3-5},\ref{3-6} (the reduced
problem describing NLE solutions). We consider a periodic
NLE solution at a certain energy $E$: $Q^{(p)}(t+2\pi / \omega_1)=Q(t)$,
$q^{(p)}_j(t + 2\pi / \omega_1) = q_j(t)$. This solution represents
a closed orbit in the phase space of the reduced problem.
If the system is integrable in a neighborhood of the closed orbit,
then each phase space trajectory from this neighborhood belongs
to some surface, which is diffeomorphic to a five-dimensional
torus \cite{via89}. Then we can introduce some new set of canonically
conjugated
coordinates $\Pi _n$, $Q_n$, $i=1,2,3,4,5$ such that
\begin{equation}
Q_n=J_n \sin{\omega_n t} \;\;, \;\;
\Pi _n = \omega_n J_n \cos{\omega_n t}\;\;. \label{a-1}
\end{equation}
Here $J_n$, $\Theta _n = \omega_n t$ denote the action-angle
variables. Without loss of generality we define the mentioned
periodic orbit with
\begin{equation}
J_1 \neq 0 \;\;, \;\; J_n = 0 \;\; (n=2,3,4,5)\;\;. \label{a-2}
\end{equation}
For trajectories from a neighborhood of the periodic orbit
the old displacements and momenta will be some functions of
the new coordinates.
These functions can be expanded in Taylor series in the new
coordinates. In the limit $J_{2,3,4,5} \ll J_1$ the old coordinates
will be a sum of the unperturbed periodic orbit solution and
four perturbations each periodic with its own frequency ($\omega_{2,3,4,5}$).
Then we can represent the perturbed solution in a Fourier series:
\begin{eqnarray}
Q(t)= \sum_{k=0,\pm 1,\pm 2,...}A_k^{(p)} {\rm e}^{ik\omega_1 t}+
\sum_{n=2,3,4,5} (A_n {\rm e}^{i\omega_n t} + A_n^* {\rm e}^{-i\omega_n t})
\;\;, \label{a-3} \\
q_j (t) = \sum_{k=0,\pm 1,\pm 2,...} a^{(p)}_{j,k} {\rm e}^{ik\omega_1 t} +
\sum_{n=2,3,4,5} (a_{j,n} {\rm e}^{i\omega_n t} + a_{j,n}^*
{\rm e}^{-i \omega_n t})\;\;. \label{a-4}
\end{eqnarray}
In the limit $J_{2,3,4,5} \ll J_1$ it follows $|A_n| \ll |A^{(p)}_{k}$
and $|a_{j,n}| \ll |a^{(p)}_{j,k}|$. Because we deal with real functions
$A^{(p)}_k=A^{(p)^*}_{-k}$ and $a^{(p)}_{j,k}=a^{(p)^*}_{j,-k}$ have
to be satisfied.

In the next step we insert \ref{a-3},\ref{a-4} into the equations
of motion of the reduced problem \ref{3-5},\ref{3-6} and set equal to
each other the terms on both sides of the five equations containing
the exponential ${\rm e}^{i\omega_n t}$. The result is a linear
set of equations for the five (infinitely small) Fourier components
$A_n,a_{j,n}$. If we set equal the terms containing the inverse exponential
the resulting algebraic equations are identical. Moreover since
the resulting set is linear and all coefficients are real,
it decomposes into two identical
sets of equations for the real and imaginary parts of the Fourier
components $A_n,a_{j,n}$. The resulting algebraic problem
is an eigenvalue problem:
\begin{equation}
\omega_n^2 \vec{r} = {\bf M}\vec{r}\;\;. \label{a-5}
\end{equation}
Here the vector $\vec{r}$ has five components and the matrix ${\bf M}$
is given by
\begin{equation}
{\bf M} =
\left(
\begin{array}{ccccc}
4C-1+\alpha & -C & -C & -C & -C  \\
-C & 4C-1+\beta & 0 & 0 & 0  \\
-C & 0 & 4C-1+\beta & 0 & 0   \\
-C & 0 & 0 & 4C-1+\beta & 0  \\
-C & 0 & 0 & 0 & 4C-1+\beta
\end{array}
\right)
\label{a-6}
\end{equation}
with
\begin{eqnarray}
\alpha = 3|A^{(p)}_0|^2 + 6\sum_{k=1,2,...}|A^{(p)}_k|^2 \;\;, \label{a-7} \\
\beta = 3|a^{(p)}_0|^2 + 6\sum_{k=1,2,...}|a^{(p)}_k|^2 \;\;. \label{a-8}
\end{eqnarray}
{}From \ref{a-7},\ref{a-8} it follows that  $\alpha=3<Q^{(p)^2}(t)>$ and
$\beta=3<q^{(p)^2}(t)>$ where the symbol $<A(t)>$ means time
average of the (periodic) function $A(t)$.
Because of \ref{a-5} all squared frequencies $\omega^2_n$ have to be
equal to the eigenvalues of the matrix ${\bf M}$ in \ref{a-6}.
Since the matrix {\bf M} has five eigenvalues, and the number of
considered frequencies was four, there is one eigenvalue left.
This eigenvalue is nothing else but the squared frequency $\omega^2_1$
of the periodic orbit itself. Indeed a small perturbation of the
given periodic orbit can be such that a new periodic orbit at
a slightly changed energy is created (because the considered
periodic orbits form a one-parameter family of solutions, where
the parameter is the frequency $\omega_1$ or the energy of the
solution). Since we exclude cases when the periodic orbit is
located on a separatrix, the change of the frequency $\omega_1$
under the considered perturbation is a smooth function of the
perturbation. Thus in first order of the perturbation (which is
considered here) the change of the frequency $\omega_1$ does not
show up (it will show up in second order of the perturbation).

The eigenvalue problem \ref{a-5},\ref{a-6} has rotational symmetry
of order 4 (this corresponds to the fact that the periodic NLE
orbit
is a symmetric solution on the squared lattice). Then there exists
a linear operator $g$ acting on the five-dimensional space spanned
by the eigenvectors $\xi_n$ of ${\bf M}$ such that
\begin{equation}
g \xi_n=\xi_{n'} \;\;, \;\; g^4 \xi_n = \xi_n \;\;. \label{a-9}
\end{equation}
It follows that there can be not more than four
assymetric eigenvectors $g \xi_n \neq \xi_n$.
But since their sum is invariant under $g$, it follows that they span
a three-dimensional subspace (cf. Appendix 10 in \cite{via89}).
Thus we end up with three assymetric
eigenvectors of \ref{a-6} with threefold degenerated eigenvalues
\begin{equation}
\omega^2_3=\omega^2_4=\omega^2_5=4C-1+\beta\;\;. \label{a-10}
\end{equation}
The remaining two symmetric eigenvectors of \ref{a-6} are
nondegenerated. They can be calculated by constructing the
$2 \times 2$ matrix of all symmetric perturbations of the
periodic orbit in analogy to the general case treated above.
The resulting frequencies are given by
\begin{equation}
\omega_{\pm}^2=\frac{1}{2}\left( \alpha + \beta \pm \sqrt{(\alpha - \beta)^2
+16C^2}\right) +4C-1 \;\;, \;\; \omega_1=\omega_-\;\;, \;\;\omega_2=\omega_+
\;\;. \label{a-11}
\end{equation}
Result \ref{a-11} can not be considered as a definition of $\omega_1$,
because one actually has to know $\alpha,\beta$ which are functions
of $\omega_1$.

In the limit of low energies (small amplitudes of oscillations) of
the reduced problem it follows $\alpha=\beta=3$ and the frequencies
will all lie inside the phonon band of the extended lattice:
\begin{eqnarray}
\omega^2_1=2+2C \;\;, \;\; \omega^2_2=2+6C \;\; , \;\; \omega^2_{3,4,5}
= 2+4C \;\;. \label{a-12}
\end{eqnarray}

\newpage

\begin{tabbing}
\large
FIGURE CAPTIONS
\normalsize
\\
\\
\\
FIG.1 \= Energy of the linearized $\Phi^4$
lattice after waiting time $t=2000$ \\
\> as a function of $\gamma_0$ (cf. text). \\
 \>Dashed line - total energy $E$ of the system; \\
 \>Solid line - energy $E_5$ stored in the central particle and \\
 \> its four neighbours. \\
\\
\\
\\
FIG.2 \> Time dependence of the energy of the linearized $\Phi^4$ lattice \\
\> for $\gamma_0 = 0.005$.  \\
\> Dashed line - total energy $E$ of the system;\\
\> Solid line - energy $E_5$. \\
\\
\\
\\
FIG.3 \> Energy distribution for the NLE solution with initial energy \\
\> $E=0.3$ after waiting time $t=3000$ (initial conditions see text). \\
\> The filled circles represent the energy values for each particle,\\
\> the solid lines are guides to the eye.     \\
\> INSERT: Time dependence of the NLE energy $e_5$.  \\
\\
\\
\\
FIG.4 \> Poincare intersection between the trajectory of the \\
\> symmetric reduced problem
 \\
\> and the subspace $\{ \dot{q},q,Q=-1, \dot Q > 0\} $. \\
\> (a) E=0.2; (b) E=0.5; (c) E=2.5; (d)=5.0. \\
\\
\\
\\
FIG.5 \> Energy dependence of the frequency.
\\
\> Filled circles - fixed point frequency from regular island Nr.1 \\
\> in Fig.4(a). \\
\> Filled squares - fixed point frequency from regular island Nr.2 \\
\> in Fig.4(a). \\
\> Open triangles - frequency $\omega_1$ of the periodic NLE solution \\
\> from the full system. \\
\> Solid line - frequency of the effective potential. \\
\> Dashed lines indicate the positions of the phonon band edges.\\
\\
\\
\\
FIG.6 \> Poincare intersection as in Fig.4(d), but:  \\
\> open circles - result for the full system; \\
\> Dots - result for the symmetric reduced problem (taken over from\\
\> Fig.4(d)). Note that because of the high density of dots\\
\> solid lines can be formed. \\
\\
\\
\\
FIG.7 \> Time dependence of the NLE energy in the full system \\
\> for initial condition corresponding to chaotic trajectory \\
\> in Fig.4(b). \\
\\
\\
\\
FIG.8 \> Poincare intersection (as in Fig.4) but for \\
\> full system with initial condition $X_{|\vec{R}|=1}=-1.0116$, \\
\> $\dot{X}_{|\vec{R}|=1}=-0.038112$, all other particles at \\
\> groundstate position and zero velocity, except velocity of \\
\> central particle $\vec{R}=0$ (adjusted to energy $E=0.3$). \\
\> The filled circles are the actual mapping results. \\
\> The lines are guides to the eye and connect the circles \\
\> in the order of their appearance. The spiral-like form \\
\> of the broken line indicates the evolution of the contraction \\
\> of the NLE solution to the fixed point periodic NLE solution.
\end{tabbing}

\end{document}